\documentstyle[prl,aps,multicol,epsfig,amsmath]{revtex}

\begin{document}
\normalsize
\draft
\widetext

\title{Theory for the ultrafast ablation of graphite films}
\author{H. O. Jeschke, M. E. Garcia and K. H. Bennemann}
\address{Institut f{\"u}r Theoretische Physik der Freien
Universit{\"a}t Berlin,
Arnimallee 14, 14195 Berlin, Germany,}
\date{\today}
\maketitle

\begin{abstract}
The physical mechanisms for damage formation in graphite films induced
by femtosecond laser pulses are analyzed using a microscopic
electronic theory. We describe the nonequilibrium dynamics of
electrons and lattice by performing molecular dynamics simulations on
time-dependent potential energy surfaces. We show that graphite has
the unique property of exhibiting two distinct laser induced
structural instabilities. For high absorbed energies ($> 3.3$~eV/atom)
we find nonequilibrium melting followed by fast evaporation. For low
intensities above the damage threshold ($> 2.0$~eV/atom) ablation
occurs via removal of intact graphite sheets.
\end{abstract}


\begin{multicols}{2}


Material irradiation with femtosecond laser pulses gives rise to many
interesting phenomena involving atomic motion, like ultrafast phase
transitions\cite{downer:92,stampfli:90,jeschke:99,soko:95,par:96,par:98,solis:98},
ablation\cite{soko:98,cola99}, optical
breakdown\cite{stuart:96,linde:96,kautek:98}, and excitation of
coherent phonons\cite{falcone:00}. Ultrafast laser ablation, in
particular, has been extensively studied mainly because of its
important technological applications\cite{krueger:99}. However, little
is known about the microscopic mechanisms leading to ablation.

We define the ablation threshold as the laser fluence for which
lattice instabilities of such magnitude are induced that the system is
irreversibly damaged and at least a monolayer of material is
removed\cite{stuart:96}. On the basis of many experimental studies it
is commonly assumed that the ablation process near the threshold is
always initiated by the melting (or ultrafast melting) of the
material.

In this letter we show that there is an exception to this general
rule. We present a theoretical study of the microscopic processes
leading to the femtosecond ablation of graphite films. We conclude
that, due to its layered, $sp^2$-bonded structure, graphite presents
the unique property of exhibiting two different ablation mechanisms
and therefore two different ablation thresholds. The mechanism for the
low fluence ablation threshold is the removal of intact graphite
sheets and does not involve melting.  In contrast, the high fluence
threshold corresponds to bond breaking processes inside the graphite
layers and leads to ultrafast melting and expansion of the film.
Moreover, we find that the metastable and short-lived low density
liquid state formed at the second ablation threshold presents some
signatures corresponding to low density liquid carbon
(LDLC)\cite{ree:99}.

We study the laser-induced ultrafast dynamics of a graphite film by
performing molecular dynamics calculations based upon a tight-binding
Hamiltonian. The effects of the strong electronic nonequilibrium
created by ultrashort laser pulses are accounted for with a method of
determining time-dependent electronic occupation numbers, which leads
to potential energy surfaces changing with time\cite{jeschke:99}. The
film geometry is simulated with a MD supercell that has periodic
boundary conditions in the horizontal directions, while in the
direction perpendicular to this, the material borders on the
vacuum~\cite{car:96}. The shape of the MD supercell is constant in
this model, and the size is sufficiently large to allow any expansion
of the material along the vertical axis. With this method we treat
the inertial confinement which is expected to be important in the
initial stages of the response of a graphite film to laser excitation.
The method of using a column-shaped MD supercell of fixed shape and
size should be a good approximation to the physical reality in the
first 10 to 100~ps after laser excitation, {\it i.$\:$e.} as long as
the vertical expansion of the material dominates and the horizontal
expansion can still be neglected.

In the MD simulations the forces acting on atom $k$ is calculated as
${\bf f}_k(\{r_{ij}(t)\},t) = - {\bf \nabla}_k\:
\Phi(\{r_{ij}(t)\},t)$, where $r_{ij} = |{\bf r}_i - {\bf r}_j|$ is
the distance between atoms $i$ and $j$, and $\Phi(\{r_{ij}(t)\},t)$ is
the potential. It is clear that, for the description of laser induced
nonequilibrium processes, $\Phi(\{r_{ij}(t)\},t)$ cannot be a model
potential, but it has to be derived from a microscopic electronic
theory, and is given by
\begin{equation}
\label{eq:phi}
  \Phi(\{r_{ij}(t)\},t) = - \sum_{m} n(\epsilon_m,t) \epsilon_m - \frac{1}{2}
  \sum_{\substack{ij \\ j\not=i}}  E_{\rm{rep}}(r_{ij}) \,.
\end{equation}
In Eq.~(\ref{eq:phi}) the quantities $\epsilon_m = \langle m|H_{\rm
TB}(\{r_{ij}(t)\}) |m \rangle$ are the eigenvalues of the electronic
Hamiltonian, $n(\epsilon_m,t)$ the corresponding electronic
occupations, and $E_{\rm{rep}}(r_{ij})$ the repulsive core-core
potential. We use a tight-binding Hamiltonian
\begin{equation}\label{eq:hamtb}
H_{\rm TB} = \sum_{i\eta} \epsilon_{i\eta} n_{i\eta} + 
\sum_{\substack{ij\eta\vartheta \\ j\not=i}}
t_{ij}^{\eta \vartheta} c_{i\eta}^+ c_{j\vartheta}^{ }\,.
\end{equation}
Here, $n_{i\eta}$ represents the occupation number operator for the
orbital $\eta$ of atom $i$, and the hopping matrix elements have been
abridged by $t_{ij}^{\eta \vartheta}$.  We follow Slater and
Koster~\cite{slater:54} in their treatment of the direction dependence
of the hopping integrals $t_{ij}^{\eta \vartheta}$. We consider the
four valence orbitals of carbon $2s$, $2p_x$, $2p_y$ and $2p_z$ (thus
$\eta,\zeta$ = $ss\sigma$, $sp\sigma$, $pp\sigma$ and $pp\pi$).  For
the distance dependence of the TB parameters $V_{ss\sigma}$,
$V_{sp\sigma}$, $V_{pp\sigma}$ and $V_{pp\pi}$, and the repulsive
potential the form proposed by Ho {\it et~al.}~\cite{ho:92} is
adopted.

Our treatment of the effect of the optically created electron-hole
plasma is founded on the following physical picture: Due to the action
of the laser pulse electrons are excited from occupied to unoccupied
levels with a time-dependent probability which is proportional to the
intensity of the laser field. As a consequence of the extremely fast
excitation process, a non-equilibrium distribution of electrons is
created. Through electron-electron collisions this electron
distribution thermalizes to an equilibrium occupation of the
electronic levels. We describe the absorption of energy by the
electronic system and its equilibration by:
\begin{equation}\begin{split}\label{eq:absorpthermal}
\frac{dn(\epsilon_m,t)}{dt} =& \int_{-\infty}^{\infty} d\omega\; 
g(\omega,t-\Delta t) 
\biggl\{\left[ n(\epsilon_m- \hbar \omega, t-\Delta t)  \right.  \\ 
& \left.  + n(\epsilon_m+ \hbar \omega, t-\Delta t)  - 2n(\epsilon_m, 
t-\Delta t)\right] \biggr\}  \\  
 & - \frac{n(\epsilon_m,t) - n^0(\epsilon_m,T_{\rm e})}{\tau_1} \,.
\end{split}\end{equation}
Thus, the electronic distribution is at each time step folded with the
pulse intensity function $g(\omega,t)$. This means that at each time
step, the occupation of an energy level $\epsilon_m$ changes in
proportion to the occupation difference with respect to levels at
$\epsilon_m - \hbar \omega$ and at $\epsilon_m + \hbar \omega$. In
Eq.~(\ref{eq:absorpthermal}), constant optical matrix elements are
assumed.  The second term of Eq.~(\ref{eq:absorpthermal}) describes
the electron-electron collisions that lead to an equilibration of the
electronic system with a rate equation of the Boltzmann type for the
distribution $n(\epsilon_m,t)$. Hence, with a time constant $\tau_1$,
the distribution $n(\epsilon_m,t)$ approaches a Fermi-Dirac
distribution $n^0(\epsilon_m,T_{\rm e})$ (with electron temperature $T_{\rm e}$).

This simple approach works very well because for dense electron-hole
plasmas extremely short relaxation times $\tau_1$ have been
found. Chemla and coworkers have reported a carrier thermalization
faster than 10~fs in GaAs~\cite{knox:88}. As we are not aware of a
measured relaxation time in graphite, we use $\tau_1=10$~fs. Note, for
such a short thermalization time the exact electronic dynamics leading
to electronic equilibrium do not play a significant role for the
structural changes we are studying here. The electronic temperature
$T_{\rm e}$ and the chemical potential $\mu$, which appear in the
Fermi-Dirac distribution and which are not determined by
Eq.~(\ref{eq:absorpthermal}), need to be fixed by an additional
principle. We demand that the nonequilibrium distribution
$n(\epsilon_m,t)$ approaches the Fermi-Dirac distribution while
conserving the total energy of the system. On the extremely short time
scale of $\tau_1=10$~fs, we expect energy loss mechanisms to be
negligible.

Using the theory described above we perform simulations on graphite
films for a wide range of pulse intensities and durations. This allows
us to determine ablation thresholds.

\begin{figure}
\begin{tabular}{cc}
\includegraphics[height=0.8\textwidth, bb=66 116 509 716, angle=-90]{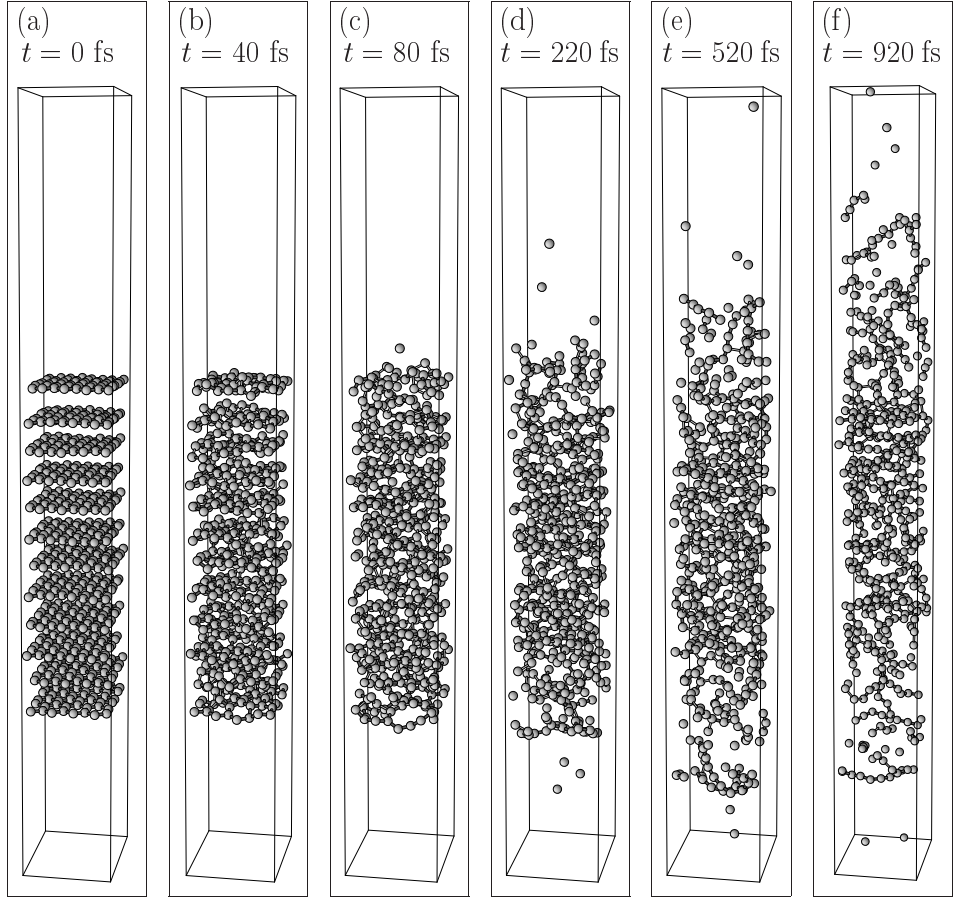}
&
\hspace*{-1.5cm}\begin{minipage}[t]{0.23\textwidth}
\caption{
Ablation of a graphite film for an absorbed energy of $E_0 =
4.0$~eV/atom. The laser pulse duration was $\tau = 20$~fs. Note the
strong expansion, the formation of a metastable liquid-like state and
the emission of carbon atoms and chains.}
\label{fig:graphabla}
\end{minipage}
\end{tabular}
\end{figure}

In Fig.~\ref{fig:graphabla} we show snapshots of a long trajectory of
a graphite film in which melting and subsequent evaporation can be
observed. The film consists of $N=576$ atoms. The dynamics shown in
this and the following figures are for laser pulses of $\tau = 20$~fs
duration with a central frequency of $\hbar\omega=3.0$~eV. We also calculated
trajectories at $\hbar\omega=2.0$~eV and found the same dynamics. The intensity
was determined so that the film absorbs $E_0 = 4.0$~eV/atom,
corresponding to a carrier density of $n_c=6.1\times 10^{22}$~cm$^{-3}$ and a
fluence of $F\approx 0.35$ J/cm$^2$\cite{fluence}. This energy density
is high enough to break the bonds of the graphite planes. At $t =
80$~fs after the pulse maximum (see Fig.~\ref{fig:graphabla}~(b)) we
see that disorder has developed inside the graphite planes and first
bonds are formed between them.  At the same time, the first carbon
monomer is evaporated from the material, confirming that the graphite
planes have been excited above their fragmentation threshold. The
following snapshots show the complete disintegration of the graphite
planes.  In Fig.~\ref{fig:graphabla}~(c) to (e), a strong volume
expansion is observed. In Fig.~\ref{fig:graphabla}~(c) to (e), four
monomers that had already moved far away from the surface were
omitted. In the advancing ablation process, further carbon monomers
are ejected, and between $\Delta t = 520$~fs and $\Delta t = 920$~fs we can
observe how the expansion of the material takes place especially
rapidly in the two surface regions of the film, leading to low
densities and the formation of short carbon chains which then start to
leave the sample.  For the ablation of graphite via the mechanism
described above we determined a threshold of
$t_{ab2}=3.3\pm0.3$~eV/atom, corresponding a fluence of $F_{ab2}\approx 0.29 \pm 0.03$
J/cm$^2$\cite{fluence}.

The examination of a graphite film which had absorbed a moderate
energy density has produced a surprising result: There is an ablation
mechanism which does not lead to a destruction of the graphite
planes. The physics of this mechanism is as follows: First, in thermal
equilibrium the graphite planes are separated by a very large distance
of $d=3.4$~{\AA} and consequently their interaction consists only of a
very weak van-der-Waals-interaction energy $E \approx
12$~meV\cite{charlier:94}. Then a photoinduced electron-hole plasma
causes a very strong vibrational excitation of the graphite layers and
at the turning point of their oscillations the atoms of two different
layers have a distance of $d \approx 2$~{{\AA}}. At this distance the
interaction becomes already quite strong and is on the order of $E \approx
1$~eV. Thus, the laser-induced strong vibrations can lead to
collisions of the planes in which momentum in the $z$ direction is
transferred. As a result, a surface plane of a graphite sample which
usually has zero total momentum can gain enough momentum to leave the
surface.

\end{multicols}

\begin{figure}
\begin{tabular}{lr}
\includegraphics[height=0.68\textwidth, bb=83 134 482 738, angle=-90]{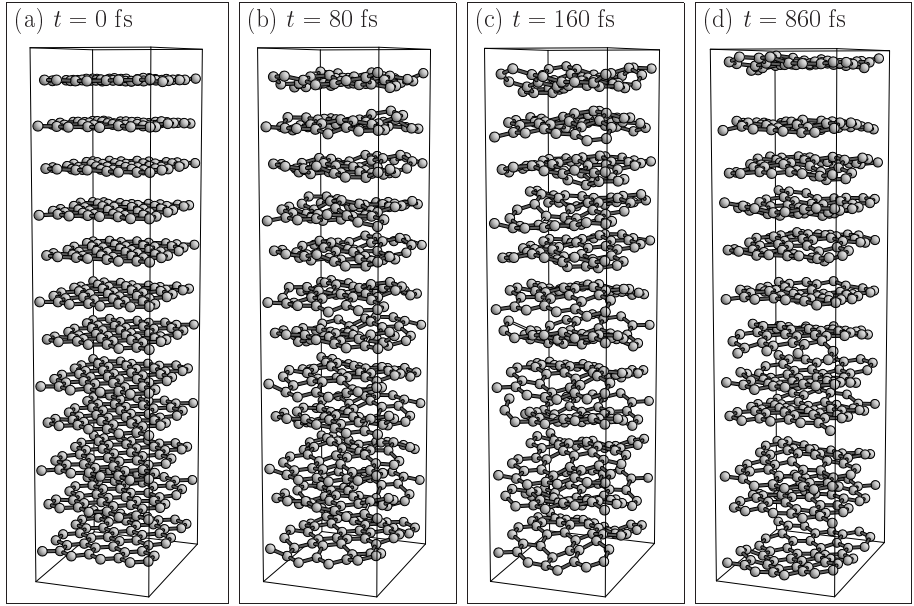}
&
\hspace*{0.5cm}\begin{minipage}[t]{0.23\textwidth}
\caption{
Ablation of graphite for an absorbed energy of $E_0 = 2.4$~eV/atom,
below the threshold for the destruction of graphite planes. The pulse
duration was $\tau = 20$~fs. Thus, the laser pulse induces strong
vibrations of the graphite planes which lead to collisions of the
planes. As a consequence, the planes at the top and at the bottom are
removed from the surface of the film.}
\label{fig:graphdrift}
\end{minipage}
\end{tabular}
\end{figure}
\begin{multicols}{2}

In Fig.~\ref{fig:graphdrift} we present snapshots of the dynamics of
graphite for a $\tau = 20$~fs laser pulse with an absorbed energy of $E_0
= 2.4$~eV/atom, corresponding to a carrier density of $n_c=3.5\times
10^{22}$~cm$^{-3}$ and a fluence of
$F\approx0.21$~J/cm$^2$\cite{fluence}. The number of atoms is $N=576$. This
trajectory shows the new ablation mechanism. At a deposited energy
density that is below the fragmentation threshold of the graphite
planes $t_{ab2}$, we find the ablation of entire graphite planes by
the mechanism explained above: The laser pulse leads to a strong
vibrational excitation of the graphite layers, with a significant
movement of the atoms perpendicular to the graphite planes. This
excitation is already strong at $t = 80$~fs after the laser pulse
maximum, but for the topmost plane it reaches a maximum at $t =160$~fs
(see Fig.~\ref{fig:graphdrift}). At that time the distance between the
graphite planes has decreased to approximately $d=1.8$~{\AA}, and the
first and second planes strongly interact. We can speak of a collision
of the two topmost graphite planes, which then leads to a momentum
transfer. Consequently, the surface plane that had up to that time
zero total momentum starts to leave the crystal in positive $z$
direction. In the same way the bottom layer of
Fig.~\ref{fig:graphdrift} can be seen to leave the rest of the film in
negative $z$ direction.

This ablation mechanism in graphite has not been described before. It
solely relies on the fact that strong vibrational excitation of the
graphite planes can increase the interaction between the planes to
such a degree that the momentum transferred in this collision process
causes ablation of entire planes. It is to be expected, of course,
that in reality we are dealing with graphite planes that show slight
defects. These would result in large plane fragments being removed
from the surface instead of entire graphite planes as the simulation
with periodic boundary conditions in the horizontal directions
suggests.  We estimate the threshold for this ablation mechanism to be
$t_{ab}= 2.0 \pm 0.4$~eV/atom\cite{tab}, corresponding a fluence of
$F_{ab}\approx 0.17 \pm 0.04$ J/cm$^2$\cite{fluence}.

The absorbed energy $E_0=2.4$~eV/atom in the example shown in
Fig.~\ref{fig:graphdrift} is high enough to cause the ejection of the
first graphite sheet within one picosecond. Note, this absorbed
energy is large enough to produce a thermal melting of the removed
graphite planes at longer times. This means that although the ablation
does not occur via melting, the ablation products might undergo
melting after some time.

\begin{figure}\hspace*{-0.5cm}
\includegraphics[width=0.47\textwidth]{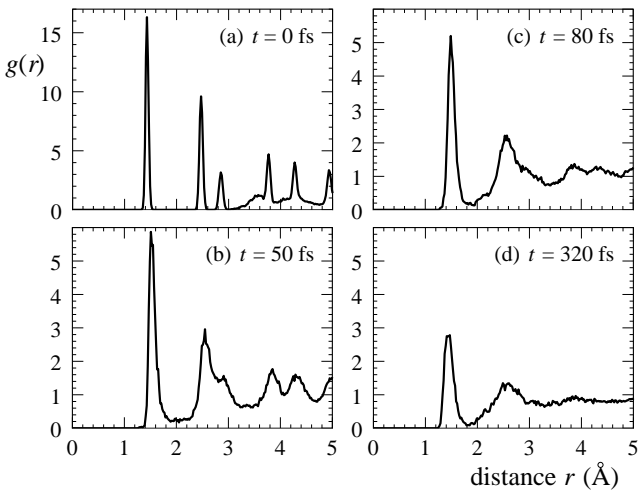}
\vspace{0.8cm}
\caption{
Pair correlation function $g(r)$ during the melting of a graphite film
for an absorbed energy of $E_0 = 4.0$~eV/atom. The pulse duration was
$\tau = 20$~fs. The pair correlation functions correspond to the same
trajectory as Fig.~\ref{fig:graphabla}. Note, the ordinate scale of
(a) is different from that of (b) to (d). }
\label{fig:graphpaircorr}
\end{figure}

We now analyze in detail the metastable liquid phase induced above
$t_{ab2}$, corresponding to the snapshots of
Fig.~\ref{fig:graphabla}. In Fig.~\ref{fig:graphpaircorr} we show the
pair correlation functions $g(r)$ during the ultrafast melting of a
graphite film. They are calculated as
\begin{equation}\label{eq:paircorr}
g(r) = \frac{\langle \Delta N(r)\rangle}{n_a \,4\pi\, r^2 \,\Delta r}\,,
\end{equation}
where $\Delta N(r)$ is the number of atoms situated at a distance between
$r$ and $r+\Delta r$ around a given particle, and $n_a=N/\Omega$ is the atomic
density of the material.  $g(r)$ at the pulse maximum (see
Fig.~\ref{fig:graphpaircorr}~(a)) does not yet differ much from $g(r)$
before the pulse. But then a rapid broadening of the structures of
$g(r)$ occurs (Fig.~\ref{fig:graphpaircorr}~(b)) and at $t=80$~fs the
pair correlation function of liquid carbon is recognizable
(Fig.~\ref{fig:graphpaircorr}~(c)). Note, because of the expansion
of the film this liquid phase has a low density and is characterized
by the presence of carbon chains. To characterize this laser induced
liquid phase we have determined the order parameter $\Psi = (\rho_4 -
\rho_2)/(\rho_4 + \rho_2)$ which was proposed by Ree {\it et al.}~\cite{ree:99}
to distinguish between the high density and the low density liquid
phase of carbon. Here, $\rho_2$ and $\rho_4$ correspond to the densities of
2-fold and 4-fold coordinated carbon atoms determined following
Brenner~\cite{brenner:90}. At a time $t=1~ps$ after the pulse maximum,
we find a value of $\Psi = -0.8$ for the graphite film shown in
Fig.~\ref{fig:graphabla} confirming that a low density, predominantly
2-fold coordinated liquid carbon phase was produced by the laser
excitation.


For a clear check of our predictions pump-probe experiments with a
time resolution of less than 100~fs and a structural resolution of a
few atomic layers would be required. Such experiments have not been
performed on graphite so far. However, the time-resolved reflectivity
measurements of laser-irradiated graphite that were reported by
Sokolowski-Tinten {\it et al.}~\cite{ultrafast:00} could be understood
qualitatively with the help of our calculations. As in the experiment
we can distinguish two different ablation thresholds
($t_{ab}$,$t_{ab2}$). Our calculated ablation thresholds are in good
agreement with the experimental ones $F_{ab} =0.185$ J/cm$^2$ and
$F_{ab2} =0.25$ J/cm$^2$ ~\cite{ultrafast:00}. This demonstrates the
validity of our approach using a tight-binding Hamiltonian. It is
important to point out that none of the existing {\it ab initio}
approaches to laser induced melting are able to take the pulse
duration and the nonequilibrium electron dynamics into account. Note
also that the number of $N = 576$ atoms in the MD supercell is not
accessible to {\it ab initio} calculations\cite{par:98,car:96}.

An important result of our approach is that both ablation thresholds
are nearly independent of the pulse duration for a range
[10fs-500fs]. For fixed pulse duration and increasing energy,
qualitative changes only occur at the ablation thresholds. Both the
emission of graphite sheets and the ultrafast melting are initiated by
strong in- and out-of-plane oscillations of the atoms.

Summarizing, the microscopic processes leading to laser ablation in
graphite films have been studied theoretically. A new ablation
mechanism has been observed at energies below the threshold for the
destruction of graphite planes. This ablation process is very
different from ablation at higher deposited energies, since large
graphite plane segments instead of small carbon clusters (chains) are
produced. The strong anisotropy of the graphite structure is crucial
for this ablation without melting. This process may play a role in the
formation of carbon nanotubes by graphite ablation.

This work has been supported by the Deutsche Forschungsgemeinschaft
through SFB 450. Our simulations were done on the CRAY T3E at
Konrad-Zuse-Zentrum f{\"u}r Informationstechnik Berlin. We acknowledge
helpful discussions with K. Sokolowski-Tinten.

\end{multicols}


\begin{references}
\bibitem{fluence}
We estimate the fluence from $F=e E_0 n_a d/(1-R-T)$, where $e$ is the
Coulomb constant, $n_a=1.76\times 10^{23}$ atoms/cm$^3$ the atomic density,
$d$ the penetration depth, $R=0.17$ the reflectivity and $T=0.65$ the
transmission\cite{downer:92}. For $d$ we assume one-photon absorption
and apply the Drude formula $d=\lambda/4\pi k$ with wave length $\lambda=620$~nm
and extinction coefficient $k=1.5$\cite{downer:92}.
\bibitem{tab} 
$t_{ab}$ was estimated taking into account also the van der Waals
interaction between the graphite sheets.
\end{references}
\end{document}